%% file: main.tex
\documentclass[submission,copyright,creativecommons]{eptcs}
\providecommand{\event}{FLACOS 2012} 

\usepackage{amsmath}
\usepackage{amssymb}
\usepackage{amsfonts}
\usepackage{stmaryrd,wasysym}
\usepackage{mathdots}
\usepackage{mathtools}

\usepackage{enumerate}
\usepackage{comment}
\usepackage{graphicx,url}
\usepackage{listings}
\usepackage[latin1]{inputenc}
\usepackage[T1]{fontenc}
\usepackage{verbatim}
\usepackage{stackrel}
\usepackage{enumitem, array,tabularx}
\usepackage{color}
\usepackage{alltt}
\usepackage{xspace}

\newcommand{\elarva}{\textsc{Elarva}\xspace}

\newcommand{\ie}{\textsl{i.e.,}\xspace}
\newcommand{\eg}{\textsl{e.g.,}\xspace}

\newcommand{\wrt}{\textsl{wrt.}\xspace}

\newcounter{Examplecount}
\title{Simplifying Contract-Violating Traces}

\author{Christian Colombo \qquad\qquad Adrian Francalanza \qquad\qquad Ian Grima
\institute{Department of Computer Science, University of Malta}
\email{\{christian.colombo $|$ adrian.francalanza $|$ igri0007\}@um.edu.mt}
}

\def\titlerunning{Simplifying Contract-Violating Traces}
\def\authorrunning{C. Colombo, A. Francalanza, and I. Grima}

\begin{document}

\maketitle

\begin{abstract}
Contract conformance is hard to determine statically, prior to the deployment of large pieces of software.   
A scalable alternative  
is to monitor for contract violations \emph{post-deployment}: once a violation is detected, the trace characterising the offending execution is analysed to pinpoint the \emph{source} of the offence.  
A major drawback with this technique is that, 
often, contract violations take time to surface, resulting in long traces that are hard to analyse.  This paper proposes a methodology together with an accompanying tool for simplifying  traces and assisting contract-violation debugging.  
\end{abstract}

\input{intro}

\input{methodology}

\input{elarva}

\input{case-study}
\input{relatedwork}

\input{concl}

\bibliographystyle{eptcs}
\bibliography{refs}

\end{document}

%% file: intro.tex
\section{Introduction}\label{sec:introduction}


Ensuring that real-world complex systems observe contract specifications  is a difficult business. Due to the large number of system states that need to be analysed, exhaustive formal techniques such as model checking are generally not feasible.   Sound static analysis techniques \cite{Nielson:2004} also suffer from these scalability issues, and often end up being too coarse, ruling out valid systems.  Testing --- a scalable solution in these cases --- is not exhaustive, thus unsound, in the sense that passing a series of tests does not imply that the contract will not be violated once the system is deployed.

\begin{figure}[!h]
\centering{\scalebox{.8}{\includegraphics{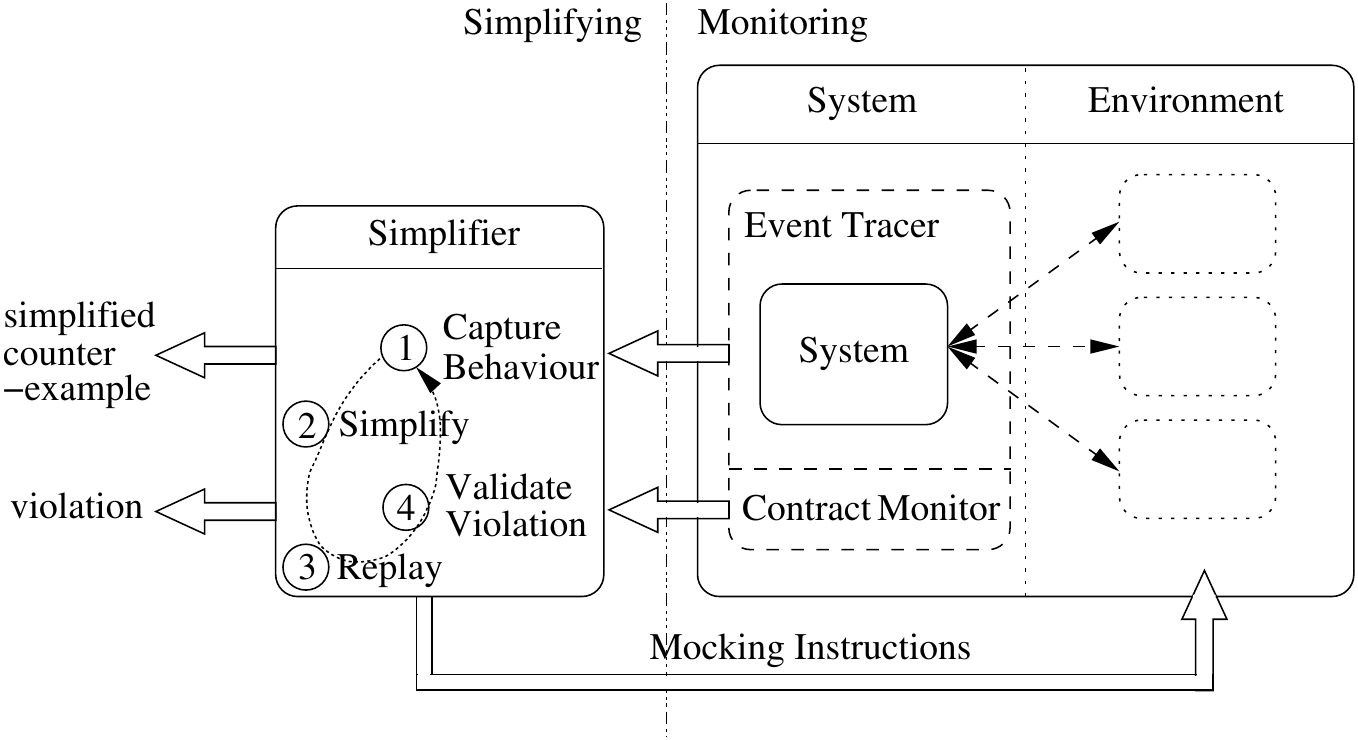}}}
\caption{Monitoring a contract (Right) enhanced with counterexample simplification (Left)}
\label{fig:elarva}
\end{figure}
     
A possible technique for dealing with this problem is to complement contract testing with the post-deployment \emph{contract monitoring} --- see Figure~\ref{fig:elarva} (Right): the contract  is synthesised as a \emph{monitor},  instrumented to run in parallel with the system (executing under an arbitrary  environment) so as to check for contract violations \emph{at runtime}.  Once a violation is detected, the monitor produces a \emph{violation trace} describing the execution that led to the contract violation and, from this trace, the cause of the violation can be inferred and rectified \emph{manually}\footnote{We are not aware of automated techniques for pinpointing the source of a contract violation from a violation-trace.}.  This technique works in principle, and can be used to prevent repeated contract-violations --- at some additional  runtime cost associated with monitor verification.  In practice, however, contract violations may be detected after a long period of monitoring, yielding violation traces that are \emph{too complex} to feasibly analyse manually.

Pinpointing the source of the problem from a violation trace can be  facilitated if  the trace is \emph{simplified}; this typically involves generating a trace  describing a shorter contract-violating execution, perhaps using simpler data values while abstracting away certain events.  This debugging aide has proved very effective in both counterexample minimisations in model checking \cite{GM07min,NSB07min} and test shrinking \cite{quickcheck,burger-issta-2011,PS11proper}. 

However, trace simplification in a post-deployment setting poses challenges.   In the absence of a system model --- as is often the case for real-world 
systems with a large number of execution states --- trace simplification typically relies  on \emph{system replaying}:  this involves re-running the offending system with simplified parameters and reduced stimuli, in the hope of obtaining a simpler execution trace that still produces the same contract violation.  But post-deployment  trace simplification based on replaying is complicated by aspects relating to system \emph{capture} and \emph{replay} \cite{burger-issta-2011,JO07scarpe,orso-woda-2006,orso05may}:            
\begin{enumerate}
\item In order to replay a system for trace-simplification analysis, one needs to infer the environment stimuli and parameters inducing the contract violation.
\item The system being monitored,  together with the environment it executes in, may be non-determi-nistic, exhibiting different behaviour under identical parameters and stimuli.
\item System replay, which may need to be carried out iteratively, 
  may produce undesirable side-effects such as writing to a database or printing on I/O terminals.
\item System replay may require  interactions with either the environment --- which cannot be controlled --- or with systems that cannot be reset (in order to recreate the same starting point).  
\item Certain computation may be too expensive and time consuming for system replay to be a viable method of finding a simpler violation trace.
\end{enumerate}

In this paper we discuss a methodology for simplifying violation traces of large 
systems in a post-deployment setting; a novel aspect of our methodology is the use of the \emph{contract information assisting the trace simplification process}.  We also present a prototype implementation of a trace-simplification tool for violating executions of Erlang programs, based on this methodology.  

The rest of the paper is structured as follows: Section~\ref{sec:meth-contr-viol} describes  our proposed methodology.  Section~\ref{sec:elarva-monit-fram} discusses an instantiation of this methodology through a tool implementation whereas Section~\ref{sec:case-study} describes a case study using this tool.  Section~\ref{sec:related-work} discusses related work and Section~\ref{sec:conclusion} concludes.

%% file: methodology.tex
\section{A Methodology for Contract-Violating Trace Simplification}
\label{sec:meth-contr-viol}

Contracts play a central role in our methodology.  Related techniques for post-deployment debugging 
fundamentally rely on \emph{terminating program executions}  (successful and not.) Contracts enables us to extend debugging to non-terminating system executions that violate a contract within a finite execution prefix; they also allowing for better separation of concerns between error definitions and system executions.

The methodology requires the contract to be  translated into an \emph{automata representation}, where transition labels correspond to actions recorded in the trace and bad states denote contract violations.\footnote{Work such as \cite{Fenech} show how contract languages such as $\mathcal{C\!\!L}$ can be automatically translated to automata representations.}     Contracts represented as automata   simplify monitor synthesis, facilitate the definition of trace simplification (see Section~\ref{sec:trace-simplification}) and  provide a lingua franca for various logics specifying contracts.


As a starting point,  our methodology  assumes the existence of two items: a violation trace and the corresponding automata-based contract that it violates. Using the mechanism depicted in Figure~\ref{fig:elarva} (Left), the methodology  uses the violated contract to guide the search for a simpler violation-trace, which constitutes the main output of the mechanism.      For this 
search, the methodology by-passes any analysis on the system source code since this analysis would not scale well for  systems of considerable size.   Moreover, the absence of a system model  forces the methodology to work with \emph{partial information}, which limits its ability to give stronger guarantees for its output; \eg it is not able to efficiently state whether the simplified trace is minimal or not, as in the case of counterexample minimisation in Model Checking \cite{GM07min,NSB07min}.  This imprecision stems from the fact that the methodology has to use \emph{simulated reruns of the system itself} as an approximating predicate for determining whether a simpler violating trace can be reached: these simulations may either not correspond to actual system executions (see Section~\ref{sec:capture}) or be fairly hard to verify because of system non-determinism (see Section~\ref{sec:replay}).     

\begin{figure}
\centering{\scalebox{.76}{\includegraphics{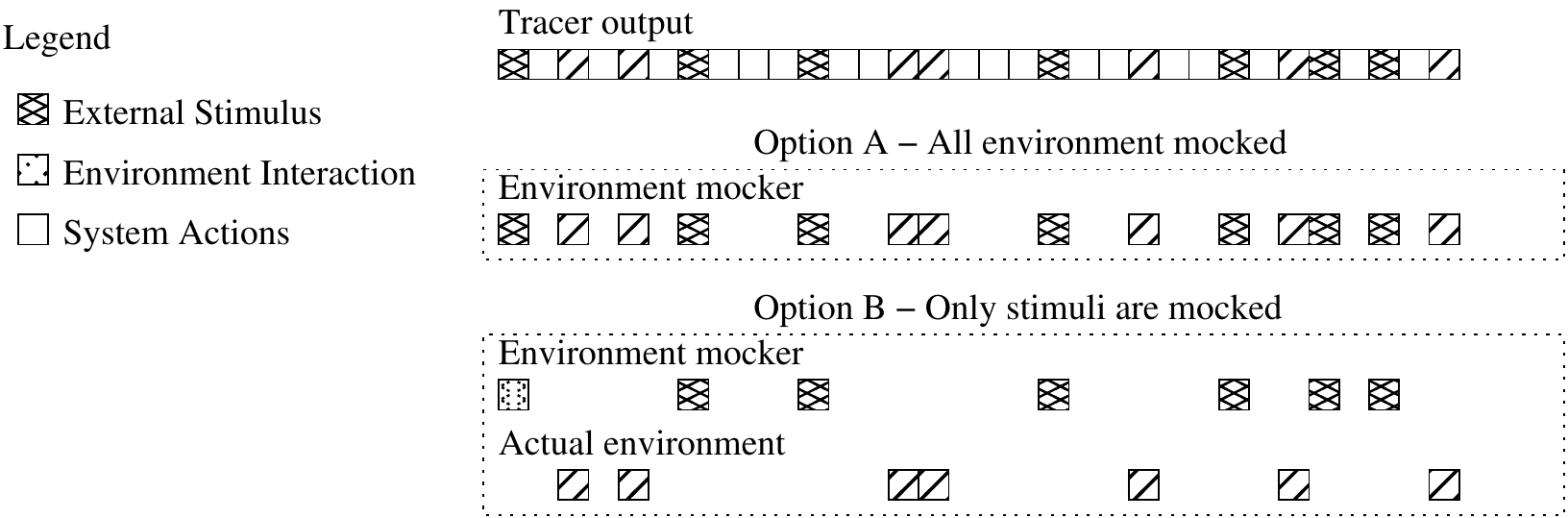}}}
\caption{Mocking the environment for replaying counterexamples}
\label{fig:mock}
\end{figure}

\subsection{Capture and Replay}
\label{sec:capture}

In order to be able to rerun the system for trace simplification purposes, the methodology needs to identify the points of interaction it had with the environment so as to be able to replicate the same execution through environment \emph{mocking}; this process is often referred to as \emph{system capture} \cite{JO07scarpe}. It involves the use of an additional subsystem during system execution, whose role is to record and identify the relevant system interaction events so that these can be later replayed (addressing complication no.\ 1 of Section~\ref{sec:introduction}).  As shown in Figure~\ref{fig:mock}, the methodology classifies events recorded  under three categories:
\textbf{External Stimuli:}
  These are computation steps \emph{instigated by the environment} on the system, that cause the system to react in certain ways.  They can range from method calls, to message sends, to spawning of sub-components in the system;     
  \textbf{Environment Interactions:} These are computation steps involving interactions between the system and the environment, that are \emph{initiated by the system}.  These include synchronous interactions such as method calls (on the environment) and returns, as well as asynchronous interactions such as communication sessions and instructions sent to the environment;
\textbf{System Actions:}  These are system computation steps that \emph{do not involve the environment}.

In a limited number of cases (\eg when the environment is resettable, deterministic and not affected by side effects) it suffices 
to mock only the environment stimuli, allowing the environment interactions to still occur with the actual environment; see Figure~\ref{fig:mock}, Option B.  However, environments rarely have these prerequisite characteristics, and 
the only other alternative is to mock \emph{both} the environment stimuli and interactions; see Figure~\ref{fig:mock}, Option A. This option carries disadvantages of its own, in that the replay may be unfaithful to the original violation execution because the environment may be \emph{stateful}: in such cases, the actual environment
would have yielded a different system interaction with a simplified trace than the mock extracted from the original violation trace.
Notwithstanding this limitation, our methodology favours Option A, because it offers better guarantees \wrt   confining the side-effects of a simulated system rerun, \ie the third complication in the list regarding post-deployment  trace simplification  of Section~\ref{sec:introduction}. 


\begin{figure}
\centering{\scalebox{.77}{\includegraphics{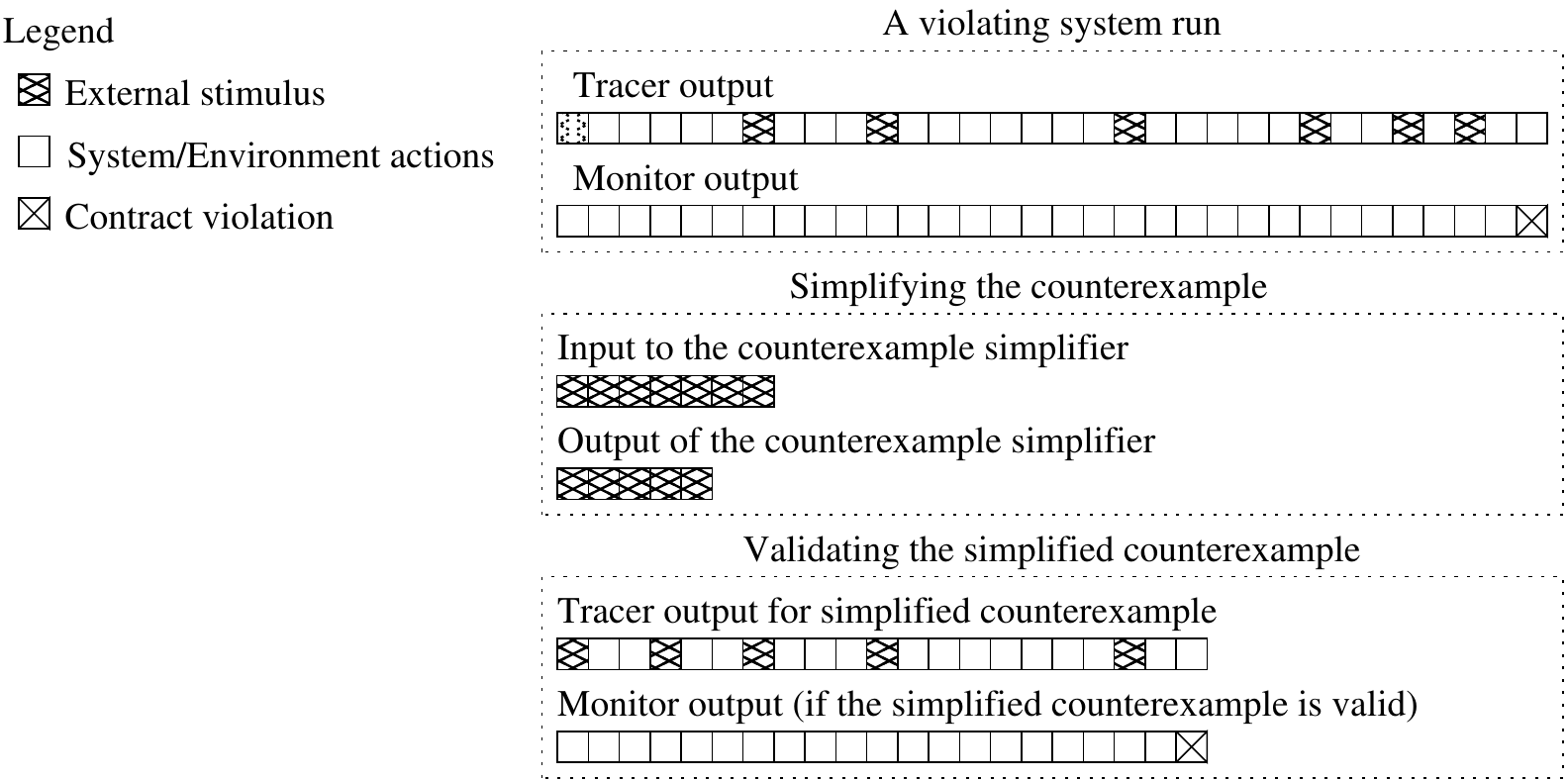}}}
\caption{Correct shrinking in terms of generated traces}
\label{fig:traces}
\end{figure}

\subsection{Trace Simplification}
\label{sec:trace-simplification}\label{sec:replay}

Trace simplification assumes the following interpretation for the \emph{simpler-than} trace relation, relying only on the structure of the trace and the contract automata: a violation trace is considered to be simpler than another violation trace whenever:
$(1)$  The violation is caused by reaching the same bad state in the contract automaton.\footnote{A contract automaton may have more than one bad state, each describing different ways how a contract may be violated.}
$(1)$  It requires less external stimuli, or the same stimuli but with simpler parameters.\footnote{This assumes some form of ordering over the data domains used.}

The underlying assumption justifying the utility of such a definition for our methodology is that traces reaching the same bad state typically relate to the same system error source. From an operational perspective, this definition also enjoys pleasing properties such as \emph{transitivity}, which adhere to intuitive notions of simplification and facilitates iterative-refinement search techniques.  Moreover, the definition integrates well with the methodology mechanism of using the captured system itself as a lightweight \emph{simpler-than} predicate: if the captured system is replayed using a \emph{subsequence} of the external stimuli of the original violation trace,  and it still violates the contract by reaching the \emph{same} bad state, then the trace generated is considered simpler.  Figure~\ref{fig:traces} depicts  this process, where the top part represents the original trace and monitor output, the middle part represents the extraction of the stimuli and the subsequent stimuli simplification (according to some criteria) while the bottom part shows the outcome of the simplification process of Figure~\ref{fig:elarva} produced from a simplified list of stimuli. 

There are other possible definitions for this relation, such as requiring that the trace is shorter in length or else that the simpler trace reaches \emph{any} bad state in the contract automata.   Despite  their respective advantages, these alternatives proved not to be as effective for our methodology.  Using the length of the trace as a measure is not compatible with iterative-refinement searching because   it yields  intermediate results that vary substantially between one another; this happens because we do not have total control over the system execution, even under capture, and  decreasing stimuli may actually result in longer traces.  On the other hand, using any bad state as a related notion of violation yielded traces that tended to describe violations caused by different sources in the system; our methodology aims to simplify the debugging for violations caused by the same source.


The methodology uses  delta-debugging techniques \cite{ZH02} to iteratively refine its search towards an improved solution:  the captured system  (which includes the synthesised monitor) is replayed under minimised stimuli and parameters as in Figure~\ref{fig:traces}. If the execution yields a violation and the violating bad state is the same as that for  the original violation trace, then the trace constitutes a simplification (as defined in Section~\ref{sec:trace-simplification}); thus the process is repeated using the minimised stimuli and  parameters of the simplified trace as the new approximates for our solution.  If not, a different minimised set of stimuli and parameters are chosen and the captured system is replayed with the new attempt.   When all minimising alternatives are exhausted without yielding a simplified trace, the current simplified approximation is returned as the output of the minimisation process. To limit the search space of simplifying traces and the computational complexity of checking whether a trace is minimal, our notion of a minimal trace is based on \emph{one-minimality} \cite{ZH02}: a trace is minimal if replaying the captured system after removing \emph{any one} of its stimuli does reproduce the bug in the original trace. 

There are however complications associated with the plain vanilla adaptation of delta debugging to our methodology.    For instance, the system itself may be non-deterministic and may yield different outcomes for the same set of stimuli and parameters --- complication no.\ 2 in the enumerated list of Section~\ref{sec:introduction};  in the absence of mechanisms forcing system replays to choose certain execution paths at non-deterministic points of execution, this can affect the precision of delta-debugging such as one-minimality guarantees \cite{ZH02}.
Our methodology tries to mitigate this imprecision by performing the same replay a number of times, using a threshold for system replays at which point the search is terminated.   

There are however  other problems.  In particular, the system may print to I/O terminals during replay --- complication no.\ 3 of  Section~\ref{sec:introduction} while  elements of the system, such as an internal database, may be stateful and not resettable to the state that lead to the original trace violation --- complication no.\ 4 of Section~\ref{sec:introduction}.  Moreover, the computational cost associated  with repeating complex system computations may make  iterative replays infeasible --- complication no.\ 5 of Section~\ref{sec:introduction}. The solution chosen by our methodology to handle these problems is to \emph{shift} the system-environment boundary we started off with in Figure~\ref{fig:elarva} (Right).  More concretely, elements of the system which produce side-effects, or are non-deterministic, can be identified and isolated (maintaining the violation), they can be considered to form part of the environment. A similar procedure can be applied to non-resettable stateful system components and computationally expensive components.  This symbolic boundary shift implies that we also \emph{mock} these components with our system capture, thus providing a system-independent, standard way of making the iterative-refinement search more precise and efficient.   \nopagebreak



%% file: elarva.tex
\section{Trace simplification for the \elarva Monitoring Framework}\label{sec:elarva-monit-fram}

We have implemented an instantiation of our methodology\footnote{The tool is freely available from \url{http://www.cs.um.edu.mt/svrg/Tools/ELARVAplus}. The distribution also includes the case study given in the paper.} as an extension to \elarva \cite{CFG11elarva}, an asynchronous monitoring tool for Erlang \cite{
  erlprog} (an actor-based programming language). 
Given the complexity of distributed industrial systems for which Erlang is usually used, contracts are a natural way how to specify what supplemented forms of behaviour the system parties are expected to adhere to. Concurrency and distribution, inherent to Erlang programs, may yield different thread interleavings each time a system is executed, potentially resulting in non-deterministic behaviour.  It is also common for Erlang systems to be programmed to execute without terminating, as in the case of controllers for network switches or  elevator systems.  These characteristics are conducive to systems with large state spaces, making  exhaustive methods of analysis infeasible.   As a result, the post-deployment setup outlined in Figure~\ref{fig:elarva},  instrumented through \elarva, constitutes  an attractive proposition for ensuring contract adherence.

In \elarva,  contracts  are specified as DATE automata \cite{CGG08FMICS} where transitions are triples of the form
\begin{verbatim}
                            event \ condition \ action
\end{verbatim}
with the following semantics: whenever an \verb+event+ occurs and, at that instant, the \verb+condition+ is satisfied, the automata transitions to the new state and the  \verb+action+ is performed.  \elarva also supports on-the-fly replication of automata through the \texttt{Foreach} construct.   More precisely, it specifies a type of contract whereby, whenever an Erlang process executing a particular function is spawned,  a corresponding  monitor executing a replica automata is launched, typically to monitor activities associated with that process.  \texttt{Foreach} constructs are particularly useful for keeping contract descriptions compact  when monitoring systems with numerous replicated processes; this is often the case for most Erlang systems where process spawning is relatively cheap \cite{erlprog}.

 \begin{figure}[!h]
 \centering \scalebox{.65} {\includegraphics{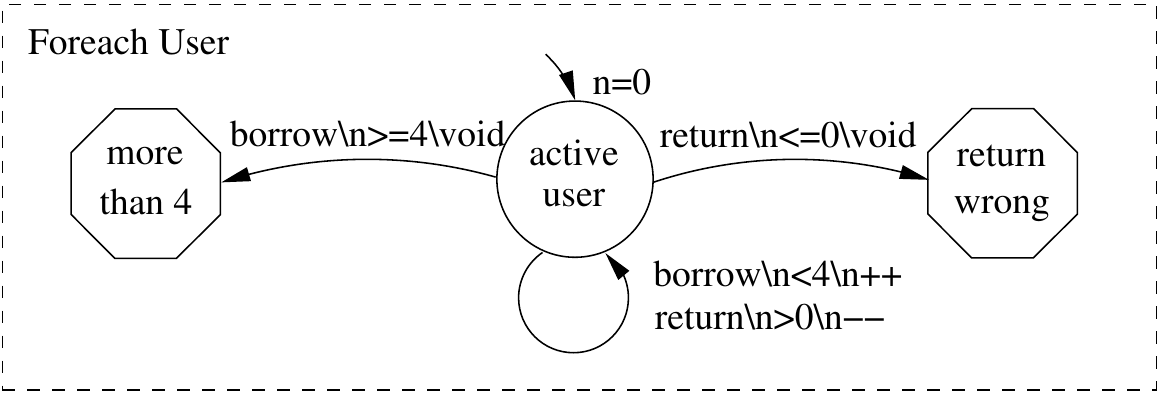}}
 \caption{Library System \texttt{Foreach} User contract}
 \label{fig:libfsm3}
 \end{figure}

Figure~\ref{fig:libfsm3} depicts an example \texttt{Foreach} contract specifying that every (process representing a) library user  can borrow a maximum of four books and, at the same time, cannot return a book when no books have been borrowed.    \texttt{Foreach}  contracts are violated if \emph{any} replicated automaton that is launched reaches the corresponding bad state (represented by octagons).


In practice, \elarva monitors Erlang programs by traversing the DATE automata in correspondence to the events read from to the program execution trace, generated by the Erlang Virtual Machine (EVM).  Erlang traces record events such as  methods calls, communication messages and process spawning: together with the type of the event and the values associated with it, Erlang traces also record the entities producing these events, \ie the unique ID of the process producing that event.    In an \elarva monitoring setup similar to that  in Figure~\ref{fig:elarva} (Right),  the EVM, acting as the Events Tracer, communicates events to the Contract Monitor while the system is executing and,  as soon as a monitor automaton reaches a bad state, it flags the violation together with the trace  justifying the violation detected.  

As  in Figure~\ref{fig:elarva} (Left), we extend the \elarva system with a \emph{Simplifier} component which takes the violation trace and the contract as inputs and produces a simplified trace as output; implicitly, the Simplifier also takes the system being monitored as input so as to carry out capture and replay. The default \elarva setting assumes that the environment consists solely of the user and, as a result,  it only mocks the user input and output interactions recorded in the trace.  This system-environment boundary may however be shifted by manually specifying the process IDs recorded in the violation trace that are to be 
 mocked.  Boundary delineation is usually a trial-and-error process; at best it can fine tune the mechanism, making the trace simplification processes more efficient and effective; at worst, no simplification is carried out \footnote{Recall that due to non-determinism the violation might not be reproduced during simplification.} and the original violation trace is returned.

\begin{figure}[h]
 \centering \scalebox{.8} {\includegraphics{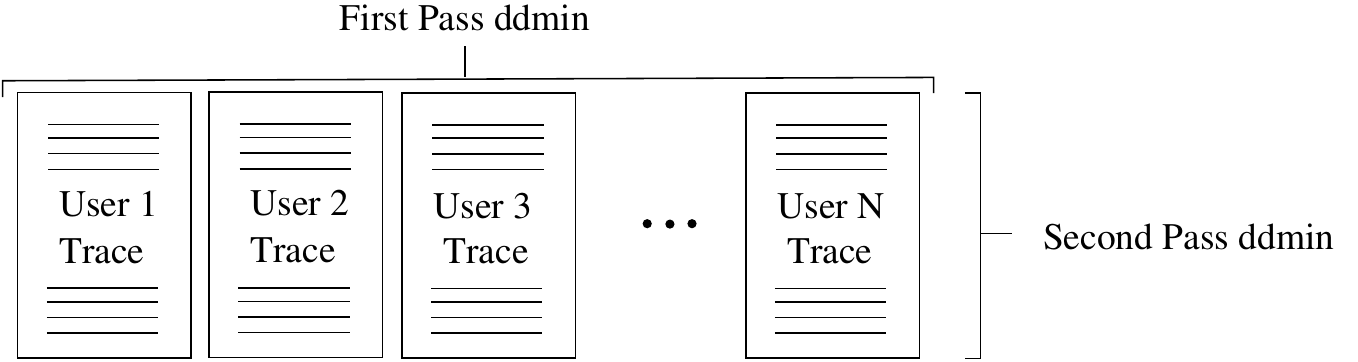}}
 \caption{Two pass \emph{ddmin} minimisation for \texttt{Foreach} Constructs}
 \label{fig:feddmin}
 \end{figure}

The  Simplifier uses an adapted version of a standard algorithm called \emph{ddmin} \cite{ZH02}; this algorithm attempts to incrementally discard parts of the trace stimuli until discarding any more  stimuli would result in a non-violating trace.\footnote{In this preliminary implementation, we do not attempt to simplify traces on the basis of simplified parameters. These techniques, used already in test-minimisation tools such as \cite{quickcheck,PS11proper} are often data-dependent and complementary to ours.}   However, in a trace with multiple stimuli, the range of possibilities can be prohibitive.  Our adaptation of the algorithm  uses the DATE automata to guide this search for stimuli minimisation.  In particular,  our heuristic is based on the assumption that a process violating a sub-property inside a \texttt{Foreach} specification would still violate it if unrelated processes executing in parallel are somehow suppressed, either through removed stimuli, or else through blocking as a result of missing environment interactions from the mocking side.    Thus, whenever the Simplifier realizes that the type of contract violated is a \texttt{Foreach} specification, it applies two passes of \emph{ddmin} trace reductions.  In the first pass, it attempts to identify which  processes correspond to different replicated instances of the same replicated automata and, subsequently attempts to incrementally suppress different groups of processes until the minimum set of groups of processes is reached that can still produce the  contract violation.  In the second pass, the Simplifier applies \emph{ddmin} again, this time on the whole trace of the remaining processes so as to further prune any stimuli which are superfluous for violating the contract.   Thus, for the example \texttt{Foreach} contract depicted in Figure~\ref{fig:libfsm3}, the Simplifier first attempts  find the minimum number of \emph{users} that can contribute to a violation  and it afterwards tries to find the minimum number of \emph{stimuli} required by this number of users leading to a violation; see Figure~\ref{fig:feddmin}.

%% file: case-study.tex
\section{Case Study}
\label{sec:case-study}

To demonstrate the effectiveness of using contract information for the violation trace simplification, we used the library case study, mentioned briefly in Section~\ref{sec:elarva-monit-fram}, which allows users to register, browse through the available books, borrow books, and eventually return the books. 
The library system should adhere to four contracts, named as follows: 
$(i)$ \emph{same book twice}: no client can borrow two books with the same name; 
$(ii)$ \emph{more than four}: no client can borrow more than four books; 
$(iii)$ \emph{different client}: no client can borrow/return a book using the name of another client; and 
$(iv)$ \emph{return wrong}: no client can return a book if  currently it is not  borrowing any. 
Encoding such contract in terms of DATEs for \elarva monitoring would result into automata such as Figure~\ref{fig:libfsm3}, which describes contracts $(ii)$ and $(iv)$ together. 
%
%
%
%

To highlight our approach we focus on contract $(iv)$ and give an example of a violation trace and how it is simplified. 
Consider the scenario where, upon starting the library system, \elarva reaches the \emph{return wrong} bad state, returning the following trace documenting the violation: 
\begin{scriptsize}
 \begin{verbatim}
[{trace_ts,<0.35.0>,'receive',{newClient,bob},{1339,842747,273000}},
{trace_ts,<0.35.0>,spawn,<0.38.0>,{client,newClient,[bob]},{1339,842747,273001}},
{trace_ts,<0.35.0>,link,<0.38.0>,{1339,842747,273002}},{trace_ts,<0.38.0>,register,bob,{1339,842747,273003}},
{trace_ts,<0.35.0>,send,{confirm_reg,bob},<0.38.0>,{1339,842747,273004}},
{trace_ts,<0.38.0>,'receive',{confirm_reg,bob},{1339,842747,273005}},
{trace_ts,<0.38.0>,send,{code_call,<0.38.0>,{ensure_loaded,client}},code_server,{1339,842747,273006}},
{trace_ts,<0.38.0>,'receive',{code_server,{module,client}},{1339,842747,273007}},
{trace_ts,<0.38.0>,send,{io_request,<0.38.0>,<0.23.0>,{put_chars,unicode,io_lib,format,[[126,110,42,
45,45,45,32,67,108,105,101,110,116,32,126,112,32,114,101,103,105,115,116,101,114,101,100,32,115,
117,99,99,101,115,115,102,117,108,108,121,32,126,110],[bob]]}},<0.23.0>,{1339,842747,273008}}, ...
\end{verbatim}
\end{scriptsize}
Apart from stimuli from the environment, the trace also contains all the messaging between the various system processes. For example while the first line is a receipt from the user to add a new client \emph{bob}, the second line is the spawn of the process which will handle \emph{bob}'s requests. Recall that since the system's internal behaviour cannot be steered, our replay and minimisation mechanisms focus on the environment stimuli. Once the trace is filtered from internal system events, the stimuli are the following: 
\begin{scriptsize}
\begin{verbatim}
[{library,{newClient,ian}},{library,{addBook,fable}}, {library,{addBook,story}},
 {library,{addBook,wish}}, {library,{newClient,bob}}, {library,{addBook,hobby}},
 {ian,{borrowBook,story}}, {ian,{borrowBook, wish}}, 	{bob,{borrowBook,fable}},
 {ian,{returnBook,story}}, {ian,{returnBook,fable}}]
\end{verbatim}
\end{scriptsize}

While the above list of stimuli is useful for reproducing the contract violation, it contains entries that do not contribute towards the violation, thus complicating debugging. The \elarva extension can apply the simplification techniques described above  and reduce the list to just two steps:  {\scriptsize\[{\verb+[{library,{newClient,bob}}, {bob,{returnBook,magic}}]+}.\] }
Note that the simplified trace obtained is in fact minimal \wrt one-minimality, \ie removing the first element would not have triggered the monitor to start checking \verb|[bob]| while removing the second element would not violate the contract. 
The simplified trace generated by the above two stimuli is what is outputted by 
\elarva, allowing the debugger to pinpoint the source of violation more easily.

\begin{table}[t]
 \centering
 \begin{tabular}{ | >{\centering\arraybackslash}m{3cm}  || >{\centering\arraybackslash}m{3cm}|| >{\centering\arraybackslash}m{1cm} |
  >{\centering\arraybackslash}m{0.9cm} ||>{\centering\arraybackslash}m{1.25cm} | >{\centering\arraybackslash}m{1.25cm} |
 }
 \hline
\small   Property Violated &  \multicolumn{1}{{>{\centering\arraybackslash}m{3cm} ||}}{ \small Original number}& \multicolumn{2}{>{\centering\arraybackslash}m{1cm} ||}{\small DDMIN}& \multicolumn{2}{>{\centering\arraybackslash}m{2.65cm} |}{\small Foreach DDMIN}\\ [0.5ex]
 
 \cline{3-6}
 \cline{3-6}
 
   & \small of Stimuli& \footnotesize Stimuli& \footnotesize Steps& \footnotesize Stimuli&\footnotesize Steps\\
 \hline\hline
  
     \footnotesize same book& \footnotesize 23 & \footnotesize 7 & \footnotesize 49 & \footnotesize 4 & \footnotesize 34 \\ \cline{2-6}
      \footnotesize twice      & \footnotesize  58    & \footnotesize 4 & \footnotesize 90 & \footnotesize 4 & \footnotesize 35 \\ \cline{1-6}
    \footnotesize more than &\footnotesize   20 &\footnotesize 15 &\footnotesize 85 &\footnotesize 10 &\footnotesize 34 \\ \cline{2-6}
    \footnotesize four &\footnotesize  73  &\footnotesize 15 &\footnotesize 279 &\footnotesize 14 &\footnotesize 70 \\ \cline{1-6}
     \footnotesize different & \footnotesize 9 &\footnotesize 2 &\footnotesize 18 &\footnotesize 2 &\footnotesize 9 \\ \cline{2-6}
   \footnotesize  client             &\footnotesize 23   &\footnotesize 2 &\footnotesize 28 &\footnotesize 2 &\footnotesize 12 \\ \cline{1-6}
   \footnotesize  return           &\footnotesize 11 &\footnotesize 2 &\footnotesize 15 &\footnotesize 2 &\footnotesize 16 \\ \cline{2-6}
   \footnotesize  wrong          &\footnotesize  60    &\footnotesize 2 &\footnotesize 35 &\footnotesize 2 &\footnotesize 23 \\ \hline 
 \end{tabular}
 \caption{Shrinking performance in different scenarios}
 \label{table:shrinkingres}
 \end{table}

To evaluate the simplification capabilities of \elarva we carried out preliminary tests on the 
the library system described above. In particular, through these tests we wanted to substantiate our hypothesis that using contract information, \ie the \emph{foreach} structure, improves the performance of the simplifying algorithm. Thus for each contract, we identified two violating traces and each trace was first simplified using the plain \emph{ddmin} algorithm and then we simplified the trace using contract information as explained in the previous section. 
The original number of stimuli, their resulting length and the number of steps required for simplification\footnote{Note that some steps were unsuccessful and did not contribute towards a simpler trace.} are shown in Table~\ref{table:shrinkingres}. 
Using contract information has consistently produced simpler traces and with the exception of one case, it has also taken fewer number of steps to reach the simplified trace.




%% file: relatedwork.tex
\section{Related Work}
\label{sec:related-work}

Two Erlang software testing tools offering trace simplification to facilitate debugging are PropEr \cite{PS11proper} and QuickCheck \cite{quickcheck}. Both tools are property-based and perform testing by randomly generating values within the given range and running the functions being tested with the generated values. Should one of the generated test cases fail, the simplification process mutates the failing test case and re-runs the result in order to check if a violation occurs. A simplified test case is considered valid if it produces an error breaking the system's specification and it is simpler than the original failing test case. These tools differ from the extended \elarva because they are used pre-deployment; since they already drive the environment, they do not need capture mechanisms.  Furthermore,  PropEr and QuickCheck do not differentiate amongst violations and a successful counterexample simplification is one which generates \emph{any} violation; by contrast, our trace simplification requires violations to match \wrt the same bad states.

In principle, our capture and replay approach is similar to SCARPE \cite{JO07scarpe,orso05may}: as we discussed in Section~\ref{sec:meth-contr-viol}, the tool leaves it up to the user to delineate the system of interest while capturing the interaction of the selected subsystem with the rest of the environment. Although  SCARPE is used for debugging purposes, it does not perform any trace shrinking.  The target language is also different from ours in that they focus on Java-based systems. 

The body of work on JINSI \cite{burger-issta-2011,orso-woda-2006} is perhaps the closest to ours. The tool is more mature than our \elarva extension and is able to use advanced  replay mechanisms and techniques such as event and dynamic slicing to considerably improve the efficiency of simplifying counterexamples.   However, their approach can only handle \emph{crashing bugs} and  \emph{terminating} program executions resulting in ``infected'' program state.\footnote{In \cite{burger-issta-2011} computation has to terminate before the state can be analysed.}  Since our work assumes the notion of a contract, it can handle  non-terminating computation as well as a notion of violations that is far richer than crashing bugs.  We also employ contract information as a form of  contract-guided event slicing, which appears to be novel.

%% file: concl.tex
\section{Conclusion}\label{sec:conclusion}

In this paper, we have studied post-deployment debugging techniques for contract violations.  Our contributions are:
$(1)$  A methodology that uses contracts to simplify violation traces that are obtained post-deployment, thus facilitating the pinpointing of the defects causing the violation.
$(2)$  A prototype tool implementation instantiating the methodology, that simplifies Erlang violation traces  using contract-based heuristics. 
When compared to state-of-the-art post-deployment debugging tools such as \cite{burger-issta-2011,orso-woda-2006}, our methodology also makes fundamental use of contracts to extend debugging beyond traces relating to  crashing bugs or terminating programs.   

\paragraph{Future Work:}
\label{sec:future-work} 
We plan to assess better and improve the simplification algorithms used by our prototype tool by exploiting more contract information such as iterating event sequences and dependencies between the events leading up to the violation. We also plan to extend our tool with parameter shrinking techniques used frequently in property based testing \cite{quickcheck,PS11proper}.  Finally, we plan to integrate thread-level scheduler tools for Erlang such as PULSE \cite{Pulse:2009} which would enable the controlled  replay of specific concurrency interleavings mirroring those of the violating trace. This approach should significantly fine-tune our capabilities of reproducing concurrency bugs.